# On the Difference Between Radio Quiet and Radio Loud AGN

P. Marziani and M. Calvani

*Osservatorio Astronomico di Padova, Italia*

J. W. Sulentic

*University of Alabama, Tuscaloosa, USA*

**Abstract.** A systematic difference in alignment between the central black hole spin and the angular momentum of the accreting gas may help explain several differences found in the optical and UV HST spectra of radio loud and radio quiet AGN.

Why do AGN appear in two distinct classes: radio loud and radio quiet? This fundamental question is basically unanswered.

A moderate amount of accreted matter leading to a total-to-initial mass ratio $M_{tot}/M_{in} \sim 2.5$ should spin up a nuclear black hole close to the maximally rotating configuration in the Kerr metric, with angular momentum per unit mass $a = 0.998$ M (Thorne 1974; $c = G = 1$). Let us assume that *all massive black holes in active nuclei at low and moderate redshift possess angular momentum*.

Elliptical galaxies are hosts of radio loud AGN. They are supported mainly by velocity dispersion, with rotation contributing to less than a few percent, and have no preferential direction for accreting gas. On the contrary, the disk plane in spiral galaxies provides a preferential plane for the accreting gas. We therefore suggest that one fundamental difference between the two classes is related to *the angle between the black hole spin and the angular momentum vector of the accreting gas*.

If the plane of the accreting gas is inclined with respect to the equatorial plane of the black hole, then the accretion flow is no longer planar. The Lense–Thirring precession should lead to an alignment of the two angular momenta close to the Kerr black hole, i.e., should twist the plane of accretion into the Kerr black hole equatorial plane. Bardeen & Petterson (1975) showed that because of viscosity the Lense–Thirring precession is important up to $\sim 10^3$ $R_g$ ($R_g = 2$ M). One half of the disk could show a concavity toward the central continuum source up to $r \sim 50000$ $R_g$ which would be visible from a distant observer. This region should be very important for optical and possibly UV line emission in AGN. We suggest a statistical difference in the distribution of warp angles between radio loud and radio quiet AGN. *In* **radio loud** *AGN the Kerr black hole is surrounded by a* **warped disk** *with large warp angle (typically $\beta_0 \approx 60°$). In* **radio quiet** *AGN the Kerr black hole is surrounded by a* **flat or nearly flat disk**.

The appeal of different warp distributions resides in offering an explanation for several statistical trends. (a) The distribution of line peak shifts for H$\beta$ obtained by Marziani et al. (1996; M96) differs significantly for radio quiet and loud AGN. Radio loud AGN show a larger spread in peak shifts (relative to the systemic velocity of the underlying galaxy), typically with $-1000$ km s$^{-1}$ $\leq \Delta v_{r,peak} \leq 1000$ km s$^{-1}$ skewed toward an excess of redshifts, and with a few outlying peaks redshifted by as much 3000 km s$^{-1}$. Radio quiet AGN show a distribution peaked at zero radial velocity with lower dispersion: 20 of 21 objects in M96 have $-400$ km s$^{-1}$ $\leq \Delta v_{r,peak} \leq 400$ km s$^{-1}$. The difference in range of $\Delta v_r$ is qualitatively consistent with our assumption on warp angle distribution. Preliminary calculations were performed assuming $r_{in} \approx 500$ $R_g$, $r_{out} \approx 50000$ $R_g$, and power law emissivity with spectral index a = 2, varying inclination from $5°$ to $35°$, and azimuthal orientation angle from $-45°$ to $45°$, and setting a fixed warp angle at $30°$. We obtained a uniform distribution covering the $\Delta v_{r,peak}$ range observed for radio loud AGN. The narrower distribution of radio quiet AGN is consistent with flatter disks. (b) Optical FeII emission is lower in radio loud AGN. As the warped region is illuminated by a larger fraction of ionizing radiation, a higher ionization parameter should induce lower FeII emission. (c) CIV$\lambda$1549 and H$\beta$ line profile parameters are correlated in radio loud AGN only; if CIV$\lambda$1549 is emitted in a wind arising from an accretion disk, and H$\beta$ from the warped part of the disk, then we may expect a correlation of line shifts and asymmetries. (d) The scatter in the luminosity correlation between H$\beta$ and CIV$\lambda$1549 is larger in radio loud AGN. We can speculate that larger warp means different obscuration for the two lines.

In addition, line emission from warped disks has the potential to explain peculiar Balmer line profiles observed in radio loud AGN that cannot be reproduced by flat disk emission, for instance (a) a profile in which the red peak is stronger than the blue one, as observed in Pictor A and 3C390.3; (b) profiles with large blue- or redshift of the line peak and extended red wings. Both species are, albeit rarely, esclusively observed in radio loud AGN.

Warped disks introduce into line profile computations the "panacea" of disk partial obscuration. Nonetheless, the additional degrees of freedom are physically grounded. The different distributions of disk warps are consistent with a bimodal distribution of radio loudness and morphology. The angular momenta orientation offers a unifying explanation that connects AGN morphology and AGN class as expected.

Remaining issues include: (a) the stability of a warped structure; (b) the failure of statistical models in explaining the predominance of shifts toward the red observed in radio loud AGN, and (c) the rather ad hoc choice of disk model parameters for peculiar line profiles such as Arp102B and OQ 208. Such outliers, however, show properties favoring ouflow ultimately related to radio loudness rather than disk warping.